%% file: main.tex
\documentclass{article}

\usepackage[square,sort,comma,numbers]{natbib}

    \usepackage[preprint]{neurips_2023}

\usepackage[utf8]{inputenc} 
\usepackage[T1]{fontenc}    
\usepackage{hyperref}       
\usepackage{url}            
\usepackage{booktabs}       
\usepackage{amsfonts}       
\usepackage{nicefrac}       
\usepackage{microtype}      
\usepackage{xcolor}         
\usepackage{graphics,graphicx} 
\usepackage[shortlabels]{enumitem}
\usepackage{graphicx,wrapfig}
\usepackage{array, makecell}%
\setcellgapes{4pt}

\usepackage{graphics,graphicx}

\usepackage[shortlabels]{enumitem}
\usepackage{algorithm}
\usepackage{algpseudocode}
\usepackage{array, makecell}%
\setcellgapes{4pt}

\usepackage{amsmath}
\algrenewcommand\algorithmicrequire{\textbf{Input:}}
\algrenewcommand\algorithmicensure{\textbf{Output:}}

\setcitestyle{square}

\title{Beyond Chemical Language: A Multimodal Approach to Enhance Molecular Property Prediction}

\author{%
  Eduardo Soares\\
  IBM Research Brazil \\
  Rio de Janeiro, RJ, Brazil \\
  \texttt{eduardo.soares@ibm.com}\\ 
  \And
  Emilio Vital Brazil \\
  IBM Research Brazil \\
  Rio de Janeiro, RJ, Brazil \\
  \texttt{evital@br.ibm.com} \\
  \And
  Karen Fiorella Aquino Gutierrez \\
  IBM Research Brazil \\
  São Paulo, SP, Brazil \\
  \texttt{fiorella.aquino@ibm.com}\\ 
  \And
  Renato Cerqueira\\
  IBM Research Brazil \\
  Rio de Janeiro, RJ, Brazil \\
  \texttt{rcerq@br.ibm.com} \\
  \And
  Dan Sanders \\
  IBM Research Almaden \\
  San Jose, CA, USA \\
  \texttt{dsand@us.ibm.com} \\
  \And
  Kristin Schmidt \\
  IBM Research Almaden \\
  San Jose, CA, USA \\
  \texttt{schmidkr@us.ibm.com} \\ 
  \And
    Dmitry Zubarev \\
  IBM Research Almaden \\
  San Jose, CA, USA \\
  \texttt{dmitry.zubarev@ibm.com}\\ 
}

\begin{document}

\maketitle

\begin{abstract}
We present a novel multimodal language model approach for predicting molecular properties by combining chemical language representation with physicochemical features.
Our approach, \textsc{MultiModal}-\textsc{MoLFormer}, utilizes a causal multi-stage feature selection method that identifies physicochemical features based on their direct causal effect on a specific target property. 
These causal features are then integrated with the vector space generated by molecular embeddings from \textsc{MoLFormer}. 
In particular, we employ Mordred descriptors as physicochemical features and identify the Markov blanket of the target property, which theoretically contains the most relevant features for accurate prediction. 
Our results demonstrate a superior performance of our proposed approach compared to existing state-of-the-art algorithms, including the chemical language-based \textsc{MoLFormer} and graph neural networks, in predicting complex tasks such as biodegradability and PFAS toxicity estimation. 
Moreover, we demonstrate the effectiveness of our feature selection method in reducing the dimensionality of the Mordred feature space while maintaining or improving the model's performance. 
Our approach opens up promising avenues for future research in molecular property prediction by harnessing the synergistic potential of both chemical language and physicochemical features, leading to enhanced performance and advancements in the field.
\end{abstract}

\section{Introduction}
Machine learning has become a popular and efficient tool for predicting molecular properties in drug discovery and material engineering \cite{butler2018machine}. 
While traditional models were trained on predefined descriptors like molecular fingerprints or geometric features \cite{pattanaik2020molecular}, recent models now focus on chemical language representations as they can facilitate the creation of molecules that exhibit specific desired characteristics \cite{born2023regression}. 
One example of chemical molecule representations are the SMILES, which is a character string representation of a molecule generated by flattening the molecular graph through a depth-first pre-order spanning tree traversal \cite{Weininger1988SMILESAC}. 
It represents important structural features of molecules, including branches, cyclic structures, and chirality information, which are not explicitly represented in the graph representation. 
However, string-based representations lack topological awareness \cite{moriwaki2018mordred}, which may hinder the ability of deep chemical language models to capture the implicit topological structure of molecular graphs \cite{shen2021out}. 

Moreover, in the domain of molecular property prediction such as the estimation of the biodegradability of organic molecules and the toxicity of per- and polyfluoroalkyl substances (PFAS), the scarcity of labeled data \cite{kirkpatrick2004chemical} is a major challenge that limits the effectiveness of supervised training for large language models \cite{wang2019smiles}. 
The cost associated with labeling molecules and the vastness of the space of plausible chemicals that require labeling further exacerbates this issue \cite{von2020retrospective}. 
Therefore, there is a growing need for molecular representation learning that can be generalized to various property prediction tasks in an unsupervised or self-supervised setting \cite{zhang2022pushing}.

Overall, chemical language-based machine learning has become a widely adopted approach for predicting molecular properties due to its efficiency and ability to accurately represent important structural features of molecules \cite{wigh2022review}. 
Recent advances in large transformers-based foundation models have demonstrated promising results in learning task-agnostic based on chemical language representations through pre-training on large unlabeled corpora and subsequent fine-tuning on downstream tasks of interest \cite{schwaller2019molecular, ross2022large}.
While pre-trained Language Models (LMs) have emerged as viable options for predicting molecular properties \cite{liu2023molrope, ross2022large, skinnider2021chemical}, they are still in their early stages of development, and there remains a need for further research to improve their performance and address limitations such as generalization and sample efficiency \cite{wang2019smiles,moret2022perplexity}.

Here we present a novel multimodal language model (\textsc{MultiModal}-\textsc{MoLFormer}) approach for predicting molecular properties, which combines chemical language representation and physicochemical features. 
Our approach employs a causal multi-stage feature selection method that selects physicochemical features based on their direct causal-effect on a specific target property to predict. 
Specifically, we use Mordred descriptors as physicochemical features and Markov blanket causal graphs as the inference algorithm to identify the most relevant features. 
Our results demonstrate that our proposed approach outperforms existing state-of-the-art algorithms, including the chemical language-based \textsc{MoLFormer} and graph neural networks, in predicting complex tasks such as the biodegradability of general compounds and PFAS toxicity estimation. 
Furthermore, we show the effectiveness of our feature selection method in reducing the dimensionality of the feature space while maintaining or improving performance.
Our approach provides a promising direction for future research in molecular property prediction, as it leverages both chemical language and physicochemical features for improved performance.

\section{Related Work}

Traditional chemical fingerprints, such as ECFP \cite{mater2019deep}, provide an efficient method to represent chemical molecules in a vector space. 
The algorithm explores the substructure connectivity around each atom of a molecule within a given radius, generating numeric representations for each substructure identified \cite{wigh2022review}. 
These representations are then combined to create a fixed-length bit string that captures the topological chemical space of the molecule.
However, while ECFP fingerprints provide useful topological information about molecules, they lack essential details about key molecular properties such as electronegativity, polarizability, and surface area \cite{moriwaki2018mordred}.
To overcome this limitation, molecular descriptors like Mordred have been developed, enabling the computation of physicochemical information about molecules \cite{keith2021combining}, which can significantly enhance the discovery process through machine learning algorithms \cite{yang2022ensemble}.

Recently, the success of language representation models in downstream natural language processing tasks has sparked interest in extending this paradigm to other domains \cite{bommasani2021opportunities}. 
Advanced neural networks, such as transformers, can be leveraged to pre-train on large unlabeled corpora and contextual language models, resulting in domain-specific ``language'' embeddings \cite{touvron2023llama}. 
This embedding becomes the exclusive input for several downstream tasks, achieving superior performance and accuracy \cite{bommasani2021opportunities}.

For example, advanced language models trained on protein sequences have been successful in understanding the language of life \cite{madani2023large}, where features extracted by the models directly from single protein sequences achieve state-of-the-art performance in downstream prediction tasks, even when used without evolutionary information \cite{ferruz2022controllable}. Similar large-scale language models pre-trained on chemical strings sequences have been explored for molecular property prediction \cite{ross2022large}. However, chemical language-based models may lack awareness of topological and physicochemical properties, which may hinder their ability to capture implicit properties of such molecules \cite{yang2022ensemble}.

To address this limitation, we propose a multimodal language model, \textsc{MultiModal}-\textsc{MoLFormer}, that uses string-based chemical representation and physicochemical features based on Mordred descriptors and selected based on their causal effect on the properties to predict. By combining these two modalities, our model can capture both the topological and physicochemical properties of molecules, improving performance in downstream prediction tasks.

\section{Overview of the proposed approach}

\begin{figure}[htbp]
    \centering
    \includegraphics[width=14cm]{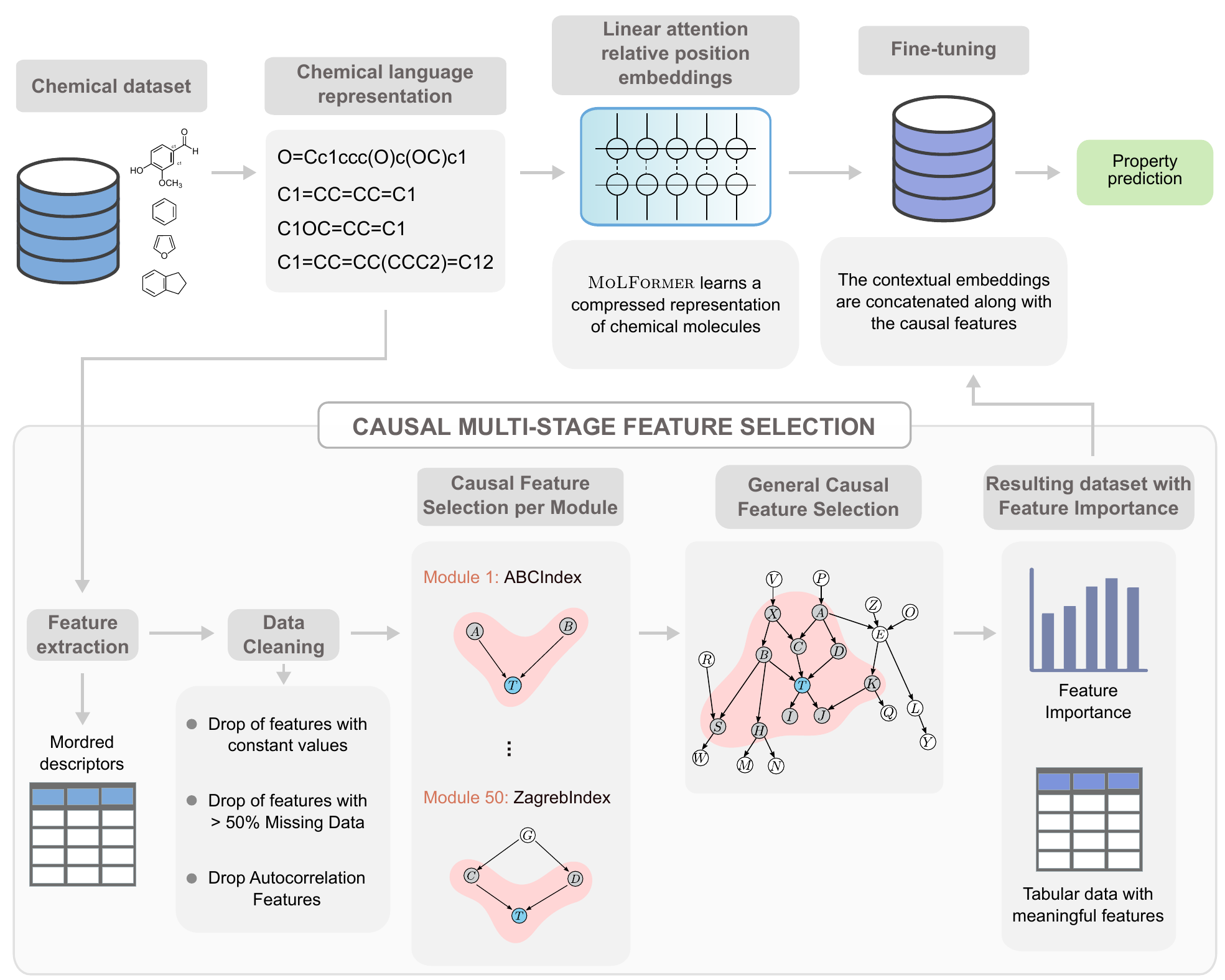}
    \caption{
The diagram above depicts the proposed architecture of \textsc{MultiModal}-\textsc{MoLFormer}, a multimodal language model designed for predicting molecular properties. This model is equipped with a causal multi-stage feature selection method, which enhances its ability to accurately predict these properties.
    }
    \label{fig:schema}
\end{figure}

The architecture of the proposed multimodal language model for molecular prediction is shown in Fig. \ref{fig:schema}. The details of the base model is described in section \ref{details}. The tokenization employed is described in section \ref{tokenization}. The multi-stage causal-based feature selection of molecular properties is presented in section \ref{Causal}. Section \ref{concatenation}, describes the concatenation process of string-based features and physicochemical features derived from the causal selection.
\label{gen_inst}

\subsection{Details of the base model}
\label{details}

Our approach is built on \textsc{MoLFormer} \cite{ross2022large}, a state-of-the-art transformer-based model for chemical language representations. \textsc{MoLFormer} is a large-scale masked language model that processes inputs through a series of blocks that alternate between self-attention and feed-forward connections. 

\textsc{MoLFormer}'s self-attention mechanism allows the network to construct complex representations that incorporate context from across the sequence. By transforming the sequence features into queries ($q$), keys ($k$), and value ($v$) representations, attention mechanisms can weigh the importance of different elements within the sequence. This enables the model to learn highly informative representations of the input data, making it a powerful tool for predicting molecular properties.

Recent studies have demonstrated that incorporating relative position embeddings between tokens results in enhanced performance \cite{liu2023molrope}. To optimize the relative encoding through position-dependent rotations $R_m$ of the query and keys at position $m$, the model uses a modified version of the RoFormer \cite{su2021roformer} attention mechanism. These rotations can be implemented efficiently as pointwise multiplications and do not significantly increase computational complexity as shown in Eq (\ref{eqz}).

\begin{equation}
Attention_m (Q,K,V)=  \frac{\sum_{n=1}^{N} \left \langle \varphi (R_mq_m), \varphi (R_nk_n)  \right \rangle v_n}{\sum_{n=1}^{N} \left \langle \varphi (R_mq_m), \varphi (R_nk_n)  \right \rangle}
\label{eqz}
\end{equation}

where $Q$,$K$,$V$ are the query, key, and value respectively, and $\varphi$ is a random feature map.

\subsection{Tokenization process and vocabulary construction}
\label{tokenization}

The approach uses a tokenization process, as detailed in \cite{schwaller2019molecular}, to create its vocabulary. This process utilizes 1.1 billion molecules from PubChem and ZINC datasets to generate 2362 unique vocabulary tokens. These tokens are then employed to fine-tune or retrain the models with a fixed embedding capacity and vocabulary size. To reduce computation time, the sequence length has been limited to a range of 1 to 202 tokens, including special tokens since over 99.4\% percent of all 1.1 billion molecules contain less than 202 tokens.

\section{Causal multi-stage feature selection}
\label{Causal}

The multi-stage feature selection method we propose follows a causal-based approach and is divided into four distinct modules (see Fig. \ref{fig:schema}). The first module is the feature extraction block, which extracts relevant features from the data. The second module is the data cleaning block.
Next, we have the causal feature selection per module block, which is the core of our method. In this module, we use Markov Blanket causal inference to identify and select features that have a causal relationship with the outcome of interest.
Finally, we have the general feature selection block, which further refines the feature selection by considering other factors such as redundancy and predictive power.

\subsection{Feature extraction}

To manipulate the chemical language representation for each molecule the RDKit \cite{landrum2013rdkit} package has been used. This allowed us to index and merge data from different sources. The RDKit was used to standardize the chemical language representations, mapping different strings belonging to the same molecule to a single molecule object.  

We then extracted molecular features using the Mordred descriptor \cite{moriwaki2018mordred}. This open-source library boasts 1826 features organized into 50 different modules, including both two- and three-dimensional descriptors.

\subsection{Data cleaning}

Following the extraction of Mordred features from the chemicals, we initiated the data cleaning process as outlined below:

\begin{enumerate}[a)]
    \item \textbf{Drop of features with constant values:} Features with constant values provide little or no relevant information to machine learning models and may introduce unnecessary noise. Thus, we removed such features during the data cleaning process.
    
    \item \textbf{Drop of features with more than 50\% of missing values:} Features with more than 50\% of missing values may not provide meaningful insights as the missing values could mask important information. 
    
    \item \textbf{Drop of Autocorrelation features:} There are more than 600 Autocorrelation descriptors. As per the findings of \cite{hollas2003analysis}, these do not correspond to structural, physical, and chemical properties of the molecule. Therefore, we opted to remove these features during the cleaning step.
\end{enumerate}

After this process, the per-module causal feature selection is initiated.

\subsection{Per-module causal feature selection}
In contrast to correlation analysis, causal-effect analysis delves deeper into the cause-and-effect relationship between a feature and a target variable \cite{guyon2007causal}. Keeping this in mind, we developed a per-module causal feature selection approach for Mordred descriptors. Since the Mordred descriptor generates over 1800 features across 50 different modules, adopting a causal approach per module proved to be an efficient strategy, optimizing computational resources.

\begin{wrapfigure}{r}{5.5 cm}
\centering
\includegraphics[width=0.96\linewidth]{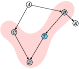}
\caption{The figure illustrate a Markov Blanket Graph where the target $T$ is direct affected by the feature $B$ and causes feature $D$.}\label{fig:MB}
\end{wrapfigure}

The Markov Blanket algorithm has been used to built the causal graphs. The Markov Blanket (MB) algorithm theoretically selects a optimal set of features considering their performance both individually and in group to predict the variable target $T$ (see \cite{koller1996toward}). Therefore, with MB($T$) it is enough to know the distribution of $T$ and all the values of the other variables that do not belong to MB($T$) become superfluous.
Next, in Fig \ref{fig:MB}, we present an example of a Markov Blanket graph that is used to represent the joint probability distribution of a set of features $\{A,B,C,D,E,T\}$ in the form of a directed acyclic graph (DAG). The nodes of the graph form the features and the directed edges form the conditional dependencies between these features. 

The MB($T$) consists of its parents (direct causes), children (direct effects), and spouses (other parents of this variable’s children) of $T$ \cite{guyon2007causal}. In the graph the variables inside the gray circle are in the MB of the target $T$ and these are used for their prediction. 

In this study, we considered the Predictive Permutation Feature Selection (PPFS) algorithm \cite{hassan2021ppfs}, which employs a Markov Blanket graph for feature selection. PPFS stands out as a versatile method since it enables subset selection for datasets encompassing both categorical and continuous features, making it applicable to both classification and regression tasks.

Markov blankets are the minimal set of features that can predict the target variable \cite{aliferis2010local}. Therefore, in the end of this process we choose the MB with the highest causal values for each module contained in the set of features descriptors. 

\subsection{General causal feature selection}
At this step, we run again the Markov Blanket based algorithm to the resulting set after joining all features descriptors selected in each module in previous step. In this sense, we filter the most important features descriptors once and minimize the redundant information between modules. 
At the end of this process, we have a parsimonious set of features with their feature importance which can be analyzed by experts. Moreover, we also have a cause-effect graph which can be used to gain deeper insights about the problem treated.

\subsection{Concatenation of chemical language-based features and physicochemical features}
\label{concatenation}
Here, we present the concatenation layer of the \textsc{MultiModal}-\textsc{MoLFormer}. At this step, the chemical language representation embeddings are combined with the physico-chemical features derived from the causal multi-stage feature selection process. 

\begin{figure}[htbp]
    \centering
    \includegraphics[width=12cm]{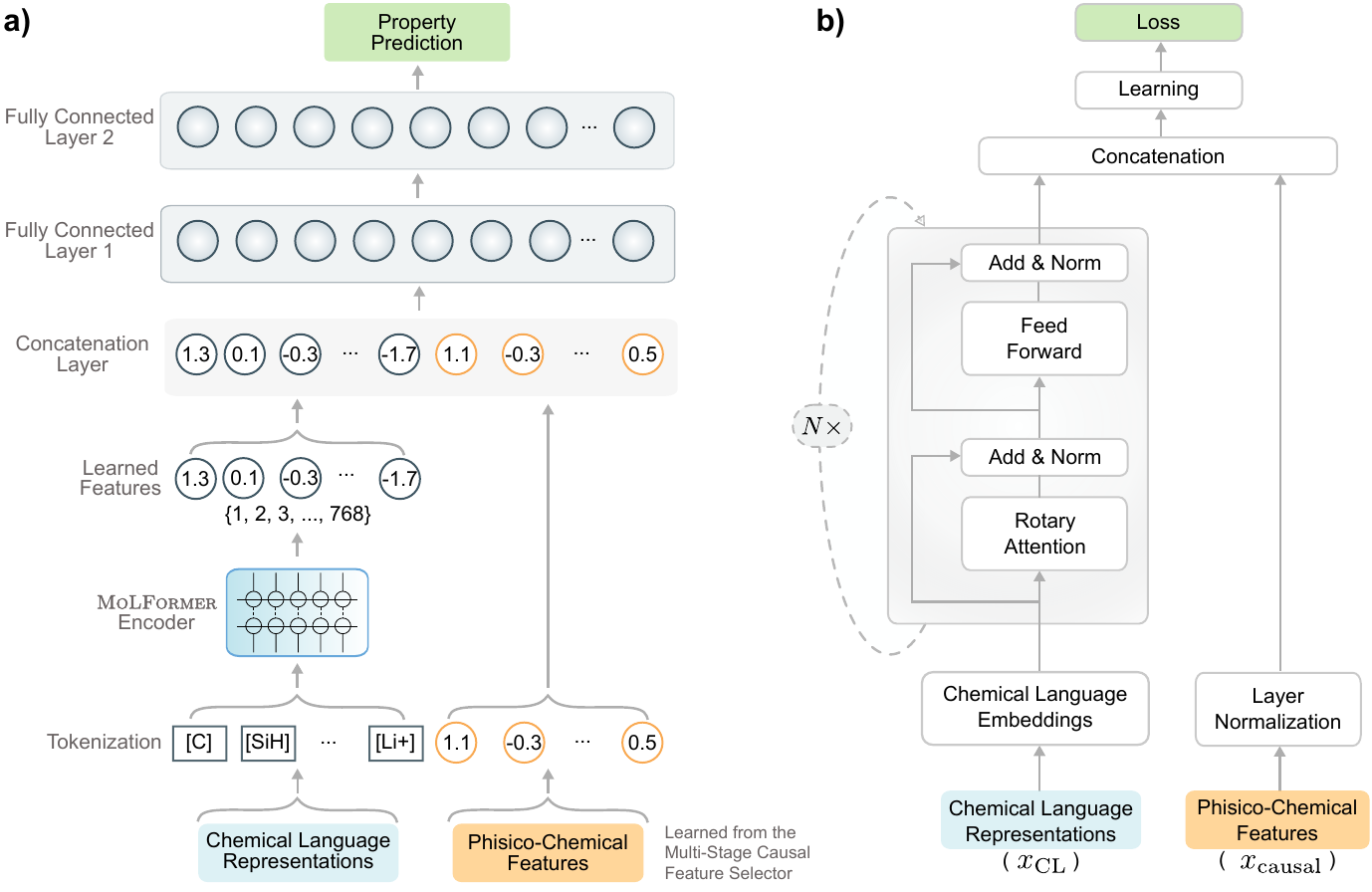}
    \caption{
        The Fig.3a illustrates the general architecture of the learning process of the \textsc{MultiModal}-\textsc{MoLFormer} with the concatenation layer. Fig.3b emphasizes the feature learning of chemical language representations. 
    }
    \label{fig:concat}
\end{figure}

Let $(x,y)$ denote a feature-target pair where $x = (x_{\text{CL}}, x_{\text{causal}})$. 
The $x_{\text{CL}}$ denotes all the features based on chemical language representations, and $x_{\text{causal}}$ refers to physicochemical features selected during the multi-stage causal-based feature selection process. Where, each $x_{\text{CL}}$ can have a sequence of tokens with length limited to a range of 1 to 202 tokens, and 768 embedding per token.  Fig. \ref{fig:concat} illustrates the concatenation architecture of the \textsc{MultiModal}-\textsc{MoLFormer} architecture.
The resulting embeddings per chemical language string, after the data transformation (Transformers layers, see Fig. \ref{fig:concat}a), is concatenated along with the physicochemical features resulting in a learning vector of dimension $(d \times e + c)$. Where, $d$ is the dimension of the dataset, $e$ is the size of the resulting embeddings and $c$ is the number of physicochemical features. The resulting vector is then passed to a learning algorithm to calculate the loss function. In this case, the learning algorithm is a Feed-Forward network with 2 fully connected layers.

\section{Experiments}

To evaluate this proposed methodology, we assessed its efficacy in tackling two demanding tasks: predicting the toxicity of PFAS substances \cite{evich2022per}, and classifying molecules by their biodegradability \cite{cheng2012silico}. 
The study of PFAS toxicity presents a captivating case, as it requires a thorough understanding of their harmful effects to effectively address and remediate their environmental impact \cite{fenton2021per}.
In the same token, the evaluation of the biodegradability of molecules offers an intriguing study case, since it is critical to establish effective assessment techniques for categorizing the possible biodegradability of organic materials \cite{su2020systemic}. 
The datasets utilized in this study are described below.

\subsection{Dataset for toxicity estimation}

To predict acute toxicity of chemical compounds we considered the LDToxDB dataset \cite{feinstein2021uncertainty} which is part of the AI4PFAS project\footnote{AI4PFAS project \url{ https://github.com/AI4PFAS/AI4PFAS.}}. 
The LDToxDB dataset is composed of 13,329 unique compounds with oral rat $\text{LD}_{50}$ aggregated from the measurements aggregated from the EPA Toxicity Estimation. Acute toxicity refers to a chemical’s propensity to cause adverse health effects within a short period.

A broader nonspecific method to measure relative toxicity of a set of compounds without any considerations of biological pathways involved is to compare median lethal doses ($\text{LD}_{50}$). 
The $\text{LD}_{50}$ metric measure the minimum dose of a compound to cause fatality in $50\%$ of laboratory subjects within 24 hours after the initial oral or dermal exposure. The Table \ref{TableRanking} contains United States Environmental Protection Agency (EPA) toxicity ranking according to the substance quantity per unit mass of laboratory-rat body weight.

\begin{table}[htbp]
\scriptsize
	\begin{center}
		\caption{EPA toxicity ranking}\label{TableRanking}
		\begin{tabular}{c|cc}
        
			\hline
            Toxicity & dosage (mg/kg body weight)& EPA Class \\
			\hline
		
			high toxicity  & $\leq 50$ &  1  \\
			
			moderate toxicity & $>50$ to 500 & 2 \\
			
			low toxicity & $>500$ to 5000 & 3 \\
			
			very low toxicity & $>5000$ & 4 \\
			
			\hline
		\end{tabular}
	\end{center}
\end{table}

\subsection{Dataset for biodegradability estimation}

In this study, we utilize the ``All-Public set'' to predict the biodegradability of compounds \cite{lee2022comparative}. The ``All-Public set'' is a composite dataset compiled from five distinct sources: I) the ECHA database; II) the NITE database; III) VEGA database; IV) the EPI Suite; V) and the OPERA suite. The dataset provides information on the biodegradability of compounds, represented by two distinct categories, ``ready-biodegradability'' (RB) and ``not ready-biodegradability'' (NRB), each containing 1097 and 1733 data samples, respectively.

\section{Results and Discussion}

The Results and Discussion section of this paper is structured into two subsections. Firstly, we detail the outcomes of our acute toxicity prediction and compare our findings with state-of-the-art methodologies. Secondly, we present the findings of our biodegradability prediction task.

\subsection{Acute toxicity prediction}

Here, we present a comparative analysis of state-of-the-art results and our proposed \textsc{MultiModal}-\textsc{MoLFormer} approach for machine learning-based $\text{LD}_{50}$ predictions using the LDToxDB dataset. Table \ref{TableTox} displays the benchmark results from the literature for state-of-the-art models, including details such as the employed machine learning methods, feature selection techniques (if applicable), input size, and performance metrics for the top-performing models. In addition to these benchmarks, we also report the results of our proposed method and the base \textsc{MoLFormer} approach.

\begin{table}[htbp]
\scriptsize
\centering
\caption{Toxicity prediction considering the LDToxDB dataset. $R^2$, $MAE$, and $ RMSE$ metrics are considered to evaluate the performance of the ML methods to predict the $\text{LD}_{50}$ dosage, \textit{acc} is used to evaluate the classification of EPA categories. For each model, the feature selection employed (if applicable) , as well as the number of input features is shown. Best results are highlighted.}
\label{TableTox}
\begin{tabular}{@{}p{2cm}p{3cm}p{2cm}p{1cm}p{1cm}p{1cm}p{1cm}@{}}
\toprule
\textbf{Method} & \textbf{Feature Selection} & \makecell{\textbf{Input Size}} & \textbf{$R^2$} & \textbf{MAE} & \textbf{RMSE} & \textbf{\textit{acc}} \\ \midrule

Multimodal-MoLFormer & \textbf{Causal Multi-stage Feature Selection} & \makecell{797} & 0.641
 & \textbf{0.277} & 0.526 & \textbf{0.84} \\
 
MoLFormer (base model)  & -- & \makecell{768} & 0.615 & 0.315 & 0.549 & 0.75 \\

XGBoost  & \textbf{Causal Multi-stage Feature Selection} &  \makecell{29} &  \textbf{0.692} & 0.334 &  \textbf{0.484} & 0.72  \\
			
Random Forest & Pearson correlation coefficient \cite{comesana2022systematic} & \makecell{106} & 0.660 & 0.337 & 0.496 & 0.70 \\
			
XGBoost  & Pearson correlation coefficient \cite{comesana2022systematic} & \makecell{106} & 0.645 & 0.360 & 0.535 & 0.71 \\
			
Random Forest  & Boruta \cite{kursa2010boruta} & \makecell{320} & 0.525 & 0.440 & 0.593 & 0.67 \\

Random Forest & RFE \cite{chen2007enhanced} & \makecell{10} & 0.233 & 0.590 & 0.791 & 0.68 \\
			
DNN \cite{feinstein2021uncertainty} & Pearson correlation coefficient & \makecell{300} & 0.658 & 0.342 &0.516 &  0.68 \\
			
GP \cite{feinstein2021uncertainty} & RF Gini & 10 Mordred / 200 ECFP bits & 0.627 & 0.376 &0.538 &  0.65 \\
			
Random Forest \cite{feinstein2021uncertainty} & -- & \makecell{1826} &  0.647 & 0.372 & 0.523 & 0.66 \\\bottomrule
\end{tabular}
\end{table}

\begin{figure}[htbp]
    \centering
    \includegraphics[width=9cm]{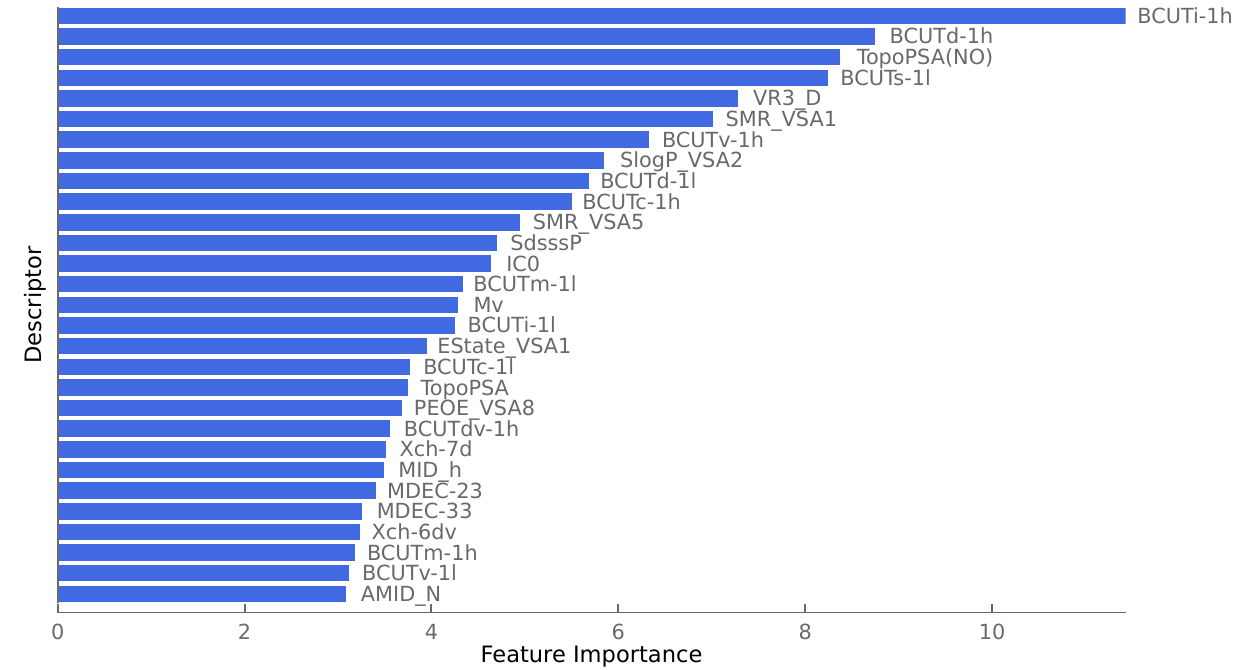}
    \caption{
        The figure illustrates the Markov Blanket feature importance histogram for the selected features in the case of compounds toxicity estimation. Note, that these scores are different from the ones given by classifiers, because these scores are given based on the causal relationship between the features and the target. Therefore, it brings pre-classification insights to experts, which is crucial to build an optimal or $quasi$-optimal dataset for a specific tasks.
    }
    \label{fig:FI}
\end{figure}

The results presented in Table~\ref{TableTox} demonstrate that the proposed multi-stage causal feature selection approach selected only 29 features that exhibit a causal relationship with the toxicity target, out of the 1826 features provided by the Mordred descriptors. 
The selected features are illustrated in Fig.~\ref{fig:FI}, which demonstrates their causal importance in predicting the toxicity of the compounds. The XGBoost model equipped with the proposed feature selection method achieved better performance than the state-of-the-art algorithms proposed in \cite{feinstein2021uncertainty}. 
This indicates the effectiveness of the proposed feature selection method in improving the performance of the model.

Incorporating the  29 physicochemical features selected during the causal multi-stage process in the \textsc{MultiModal}-\textsc{MoLFormer} model resulted in a significant improvement in the classification accuracy for EPA categories, from $0.75$ to $0.84$, when compared to the base \textsc{MoLFormer} approach. While the \textsc{MoLFormer} base model has the ability to learn from chemical language representations, including task-specific physicochemical features can enhance the model's understanding and performance for specific tasks, such as toxicity estimation.

It is worth noting that the $RMSE$ and $R^2$ metrics suggest that the \textsc{MultiModal}-\textsc{MoLFormer} model may miss outliers values, but the MAE metric indicates that the model performs well on average, leading to better classification accuracy. Although further investigation is required to address the potential limitation of the model in handling outliers, our results suggest that incorporating physicochemical features can improve the overall performance of the \textsc{MultiModal}-\textsc{MoLFormer} model for predicting molecular properties.

Overall, these results demonstrate the effectiveness of the proposed \textsc{MultiModal}-\textsc{MoLFormer} approach for predicting the toxicity of compounds, and the importance of incorporating physicochemical features in improving the model's performance.

\subsection{Biodegradability classification}

The biodegradability classification task results are presented in Table \ref{TableBio}. This task aims to classify compounds into one of two classes: ``ready-biodegradability'' (RB) or ``not ready-biodegradability'' (NRB). The RB class denotes compounds that are expected to be readily biodegradable in the environment, while the NRB class denotes compounds that are not. The table displays the performance metrics for the state-of-the-art machine learning models evaluated on the ``All-Public set'' dataset. Specifically, the Table \ref{TableBio} includes the name of each model, the feature selection method (if any), the input size, and the evaluation metrics.  

\begin{table}[htbp]
\scriptsize
\centering
\caption{Biodegradability classification considering the ``All-Public set''. \textit{acc}, \textit{Specificity}, and \textit{Sensitivity} metrics are considered to evaluate the performance of the studied methods to evaluate the classification of biodegradability classes. Best results are highlighted.
}
\label{TableBio}
\begin{tabular}{@{}p{2cm}p{3cm}p{2cm}p{1cm}p{2cm}p{2cm}@{}}
\toprule
\textbf{Method} & \textbf{Feature Selection} & \textbf{Input Size} & \textbf{\textit{acc}} & \textbf{\textit{Specificity}} & \textbf{\textit{Sensitivity}}  \\ \midrule

Multimodal-MoLFormer & \textbf{Causal Multi-stage Feature Selection} & \makecell{773} & \textbf{0.94}
 & \textbf{0.939} & \textbf{0.942}  \\
 
MoLFormer  (base model)  & \makecell{--} & \makecell{768} & 0.84 & 0.81 & 0.87  \\

XGBoost  & \textbf{Causal Multi-stage Feature Selection} &  \makecell{\textbf{5}} &  0.86 & 0.85 & 0.86  \\
			
$k$NN \cite{lee2022comparative} & Genetic Algorithm& \makecell{642}  & 0.83 & 0.85 & 0.81\\
			
SVM \cite{lee2022comparative} & Genetic Algorithm& \makecell{50} & 0.84 & 0.82 & 0.86 \\
			
Random Forest \cite{lee2022comparative}&  Genetic Algorithm & \makecell{832} & 0.84 & 0.81 & 0.86  \\

Gradient Boosting \cite{lee2022comparative} & Genetic Algorithm & \makecell{470} & 0.84 & 0.78 & 0.90  \\\bottomrule
\end{tabular}
\end{table}

The experimental results demonstrate the effectiveness of the proposed approach for biodegradability classification, surpassing the performance of previous benchmark models in terms of accuracy, precision, and sensitivity. This is a significant contribution to the development of sustainable and environmentally friendly materials, as the ability to accurately classify compounds based on their biodegradability potential can help in the design and selection of materials with minimal environmental impact.

\begin{figure}%
    \centering
    \includegraphics[width=8cm]{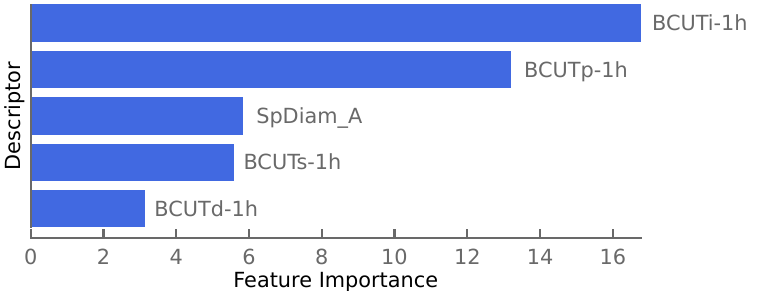} %
    \qquad
    \includegraphics[width=4cm]{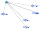}%
    \caption{The figure illustrates the Markov Blanket feature importance and graph given by the causality feature selection process.}%
    \label{fig:FIbiodeg}%
\end{figure}

Interestingly, the proposed causal feature selection approach was able to achieve better results with just 5 features (Fig. \ref{fig:FIbiodeg}), compared to the more than 50 features used by the state-of-the-art models to obtain a classification accuracy of 84\%. This not only improves the interpretability of the models, but also provides experts with greater insights into the underlying causal mechanisms involved in biodegradability classification. Moreover, it enhances the performance of \textsc{MultiModal}-\textsc{MoLFormer} compared to the base \textsc{MoLFormer} model, by aggregating features with relevant causal effects over the biodegradability classes.

These results highlight the importance of considering causal relationships between the features and the target variable in machine learning applications, particularly in domains where interpretability and explainability are critical. The proposed approach provides a promising direction for future research in the field of biodegradability classification and other related environmental applications.

In both experiments, we could observe that the proposed feature select approach helped to improve the metrics over these challenging tasks. We could note also, an improvement in terms of model's compactness, which is useful to experts to understand the process, and obtain deeper insights about such tasks. Moreover, the method proposed here provides feature importance about the causal effect of selected features over the target, which is helpful when building a dataset for a such specific task.

\section{Conclusion}

In this paper, we have introduced a novel multimodal approach for predicting molecular properties that leverages both chemical language representations and physicochemical features. Our proposed approach incorporates a unique causal multi-stage feature selection method that selects relevant physicochemical features based on their direct causal-effect on a specific target property to predict. Using Mordred descriptors as physicochemical features and identifying the Markov blanket of the target property for feature selection, our approach has demonstrated superior performance in predicting complex tasks such as general compounds biodegradability and PFAS toxicity estimation.

The effectiveness of our feature selection method in reducing the dimensionality of the Mordred feature space while maintaining or improving performance is noteworthy. Our proposed approach has outperformed existing state-of-the-art algorithms, including the chemical language-based only \textsc{MoLFormer} and graph neural networks. Although further investigation is needed to address the potential limitation of the model in handling outliers, our results suggest that incorporating physicochemical features can improve the overall performance of the \textsc{MultiModal}-\textsc{MoLFormer} model for predicting molecular properties.

These findings offer a promising direction for enhancing the accuracy and effectiveness of molecular property prediction. By incorporating both chemical language and physicochemical features, our approach has the potential to make significant contributions to drug discovery, materials science, and other fields that depend on accurate prediction of molecular properties. Future research may build upon our approach to further improve molecular property prediction and advance the understanding and application of chemical language and physicochemical features.

\footnotesize
\bibliographystyle{IEEEtran}
\bibliography{sample_FM}

\include{supplementary_material}


\end{document}

%% file: supplementary_material.tex
\section*{Supplementary Material}
For both experiments, toxicity estimation of PFAS molecules \cite{feinstein2021uncertainty} \footnote{AI4PFAS project \url{ https://github.com/AI4PFAS/AI4PFAS.}} and biodegradability of general molecules \cite{lee2022comparative}, we divided the datasets into: 60\% for training purposes, 20\% for validation purposes, and 20\% for testing purposes. In both cases, the divisions have been made following the original papers in order to have a fair comparison with the state-of-the-art algorithms.
Table \ref{TableTox} refers to the hyper-parameters considered for the toxicity estimation task. 

\begin{table}[htbp]
\scriptsize
\centering
\caption{\textsc{MultiModal}-\textsc{MoLFormer} hyper-parameters for toxicity prediction considering the LDToxDB dataset.}

\label{TableTox}
\begin{tabular}{@{}p{4cm}p{2cm}}
\toprule
\textbf{Hyper-parameter} & \textbf{Value}  \\ \midrule

Device & cuda  \\
GPU & 1 \\
Batch Size & 32  \\
Number of heads & 12  \\
Number of embeddings & 768  \\
Dropout & 0.1  \\
Learning rate & 3e-5\\
Number of workers & 8\\
Number of epochs & 2000\\
Dimension of the Feed-Forward network & [797 797 797 1] \\
 
 \\\bottomrule
\end{tabular}
\end{table}

Table \ref{TableBiodeg} refers to the hyper-parameters considered for biodegradability classification..

\begin{table}[htbp]
\scriptsize
\centering
\caption{\textsc{MultiModal}-\textsc{MoLFormer} hyper-parameters for biodegradability prediction.}

\label{TableBiodeg}
\begin{tabular}{@{}p{4cm}p{2cm}}
\toprule
\textbf{Hyper-parameter} & \textbf{Value}  \\ \midrule

Device & cuda  \\
GPU & 1 \\
Batch Size & 32  \\
Number of heads & 12  \\
Number of layers & 12  \\
Number of embeddings & 768  \\
Dropout & 0.1 \\
Learning rate & 3e-5\\
Number of workers & 8\\
Number of epochs & 2000\\
Dimension of the Feed-Forward network & [773 773 773 1]\\

\\\bottomrule
\end{tabular}
\end{table}

Both experiments have been conducted in the IBM computing cluster with $32$ GB of RAM memory and a  Nvidia V$100$ GPU.

\begin{figure}[htbp]
    \centering
    \includegraphics[width=12cm]{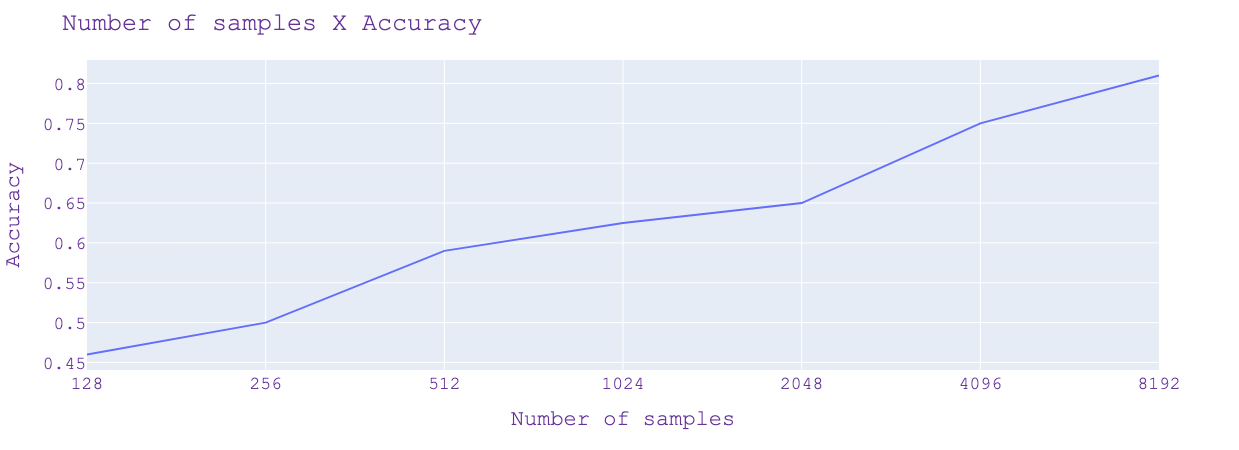}
    \caption{
        This figure illustrates the \textsc{MultiModal}-\textsc{MoLFormer} learning evolution with different training data samples sizes. 
    }
    \label{fig:evol}
\end{figure}

Fig. \ref{fig:evol} illustrates the evolution of the \textsc{MultiModal}-\textsc{MoLFormer} in terms of classification accuracy for the toxicity estimation task considering different numbers of training samples. Through this analysis, we could observe that the proposed \textsc{MultiModal}-\textsc{MoLFormer} needed $4096$ samples to obtain similar classification accuracy to the original \textsc{MoLFormer} approach that used the full training set. This indicate us, the influence of the physico-chemical features to improve the learning of the proposed approach, even considering less training data samples. 

The experiments conducted in this research are based on the recently introduced \textsc{MoLFormer} approach which can be accessed through this link (\url{https://github.com/IBM/molformer}). 

To extract the physico-chemical features based on the Mordred descriptors we used the library available at \url{http://mordred-descriptor.github.io/documentation/v0.1.0/introduction.html} \cite{moriwaki2018mordred}. To build the Markov-Blanket causal feature selection blocks we used the library available at \url{https://github.com/atif-hassan/PyImpetus} \cite{hassan2021ppfs}. The default parameters were used for the physico-chemical extraction from SMILES and to build the Markov-Blanket causal graphs.